\documentclass[11pt,bibliography=totoc]{scrartcl}

\usepackage{soul}
\usepackage[utf8]{inputenc}
\usepackage[english]{babel}
\usepackage{graphicx,amsmath,amsfonts,amssymb,dcolumn,amsthm,slashed,bbm,xspace,pgffor,tikz,mathbbol,latexsym,cite,braket}
\usepackage{microtype}
\usepackage{lmodern}
\usepackage[normalem]{ulem}
\usepackage{xcolor}
\usepackage[colorlinks=true,citecolor=blue,urlcolor=blue,linkcolor=blue]{hyperref}
\usepackage[hang]{footmisc} 
\usepackage[affil-it]{authblk}
\usepackage{multicol}

\numberwithin{equation}{section}

\begin{document}

\title{\textbf{ Configuration entropy in the soft wall AdS/QCD model  and the   Wien law}}

\author{Nelson R.~F.~Braga$^{a}$\thanks{\href{mailto:nrfbraga@gmail.com}{nrfbraga@gmail.com}},~  Octavio C.~Junqueira$^a$\thanks{\href{mailto:octavioj@pos.if.ufrj.br}{octavioj@pos.if.ufrj.br}} }
\affil{\footnotesize $^{a}$ UFRJ --- Universidade Federal do Rio de Janeiro, Instituto de Física,\\
Caixa Postal 68528, Rio de Janeiro, Brasil}

\date{}
\maketitle

\begin{abstract}
The soft wall AdS/QCD holographic model provides simple estimates for the spectra of  light mesons and glueballs satisfying linear Regge trajectories. It is also an interesting tool to represent the confinement/deconfinement transition of a gauge theory, that is pictured as  a Hawking-Page  transition  from a dual geometry with no horizon to a black hole space. A very interesting tool to analyze stability of general physical systems  that are localized in space is the configuration (or complexity)  entropy (CE). This quantity, inspired in Shannon information entropy, is defined in terms of the energy density of the system in momentum space. The purpose of this work is to use the CE  to investigate the stability of the soft wall background as a function of the temperature. A nontrivial aspect is that the geometry is an anti-de Sitter black hole, that has a singular energy density. In order to make it possible to calculate the CE, we first propose a regularized form  for the  black hole energy density. Then, calculating the CE, it is observed  that its  behavior is consistently related to the black hole instability in anti-de Sitter space.    Another  interesting result that emerges from this  analysis is that the regularized   energy density shows a behavior similar to the Wien law, satisfied by black body radiation. That means: the momentum $ k_{max} $ where the energy density  is maximum, varies  with the temperature $T$  obeying  the relation: $  T / k_{max} =  constant $ in the deconfined phase.

\end{abstract}

\section{Introduction}
It was proposed in \cite{Gleiser:2011di,Gleiser:2012tu,Gleiser:2013mga}  that the configuration entropy (CE)  works  as an indicator of  the stability of physical systems. The CE, also known as complexity entropy, is motivated by information theory, in particular by the Shannon infomation entropy \cite{shannon}. Currently, one finds in the literature many interesting examples  where the CE plays the role of representing stability, as for instance  \cite{Bernardini:2016qit,Braga:2017fsb,Braga:2018fyc,Ferreira:2019inu,Braga:2019jqg, Correa:2015vka,Braga:2016wzx,Karapetyan:2016fai,Karapetyan:2017edu,Karapetyan:2018oye,Karapetyan:2018yhm,Lee:2018zmp,Bazeia:2018uyg,Ma:2018wtw,Zhao:2019xle,Ferreira:2019nkz,Karapetyan:2019ran,Braga:2020myi,Karapetyan:2020yhs,Ferreira:2020iry,Alves:2020cmr,MarinhoRodrigues:2020ssq,Stephens:2019tav,Bazeia:2021stz,Braga:2020opg}. By comparing the CE for different states of the same system, it was observed, for diverse   physical systems, that the smaller is the CE, the more stable is the state.

The Shannon information entropy \cite{shannon} for a discrete variable with probabilities $p_n$  for the possible values that it can assume is defined as:
 \begin{equation}
 S = - \sum_n  \, p_n  \log p_n \,.
 \label{discretepositionentropy}
 \end{equation}
 It is a measure of the information contained in the variable. The configuration entropy is defined as a continuous version of \eqref{discretepositionentropy}, that for a one-dimensional system reads   
  \begin{equation}
 S_C[f]  = - \int dk  \, f({k }) \ln f({k }) \,,
 \label{positionentropy}
 \end{equation}
 The quantity $ f (k) $ is called modal fraction and is usually defined in terms of the energy density in momentum space, $ \rho (k) $, namely,   
 \begin{equation}
  f(k) = \frac{\vert  \rho (k)\vert^2}{\vert  \rho(k)\vert^2_{max}  }
 \label{modalfraction}
 \end{equation}
 where $  \vert  \rho(k)\vert^2_{max} $ is the maximum value assumed by  $ \vert  \rho (k)\vert^2 $. The modal fraction can alternatively be defined to be a normalized function, replacing in the denominator of eq. (\ref{modalfraction}) the square of the maximum value of the energy density in momentum space by $\int \vert  \rho (k)\vert^2  dk $. Such a definition would be more similar to the Shannon entropy, where the probabilities are normalized, but lead to negative values for the CE. This happens because, in contrast to    (\ref{discretepositionentropy}) where the probabilities satisfy $p_n \le 1$, the same rule does not apply to the continuous case, that involves densities.  
  
  There are many situations in physics where is important to understand what are the conditions that affect the stability of a system. In this work we are interested in the case of the holographic soft wall model\cite{Karch:2006pv}. This  AdS/QCD holographic model provides simple estimates for meson and glueball masses and also for the confinement/deconfinement temperature \cite{Herzog:2006ra,BallonBayona:2007vp}.  The model assumes an approximate duality beween fields in an anti-de Sitter (AdS)   space with a scalar background and light mesons in a strongly coupled gauge theory.  In the finite temperature case, the geometry changes. For   temperatures $ T $ above a critical value $  T_c$ the geometry dual to the gauge theory is an AdS black hole, while for  $ T < T_c$  it is just a thermal AdS space. The black hole phase corresponds to the deconfined plasma phase of the  gauge theory. Stability of the black hole  space corresponds, in this case,  to stability of the plasma.   The confinement/deconfinement transition is represented in holography in  terms of a Hawking-Page (HP) transition \cite{Hawking:1982dh}, as pointed out by Witten   \cite{Witten:1998zw} .

 The main  purpose of the present work  is to analyse the stability of the soft wall model using the configuration entropy approach. As we will show, the energy density is singular, so it is necessary to develop a regularization procedure in order to find a density that can be used to define the modal fraction and then the CE. A remarkable property of this regularized density is that it obeys a relation similar to the Wien law for black body radiation. The maximum value of the energy density occurs in a momentum $ k_{max} $  that is proportional to the temperature. The organization is the following. In section 2 we present a review of the soft wall model at finite temperature and explain how does the Hawking-Page transition shows up. Then in section 3 we develop a regularized form for the energy density. Section 4 is devoted to discuss the behaviour similar to the Wien law displayed by such density. Finally, in section 5 we present the result for the CE of the soft wall as a function of the temperature. A discussion about the results and some conclusions are presented in section 6.

\section{Soft wall AdS/QCD model and Hawking-Page transition}

In the soft wall AdS/QCD model \cite{Karch:2006pv}, the gravitational part of the action  is given by
\begin{equation}
I = -\frac{1}{2 \kappa^2} \int d^5x \sqrt{g}e^{-\Phi} \left( R + \frac{12}{L^2}\right)\;,
\end{equation}  
where $\kappa$ is the gravitational coupling and $R$  the Ricci scalar. The   dilaton field  $\Phi = cz^2$ plays the role of introducing, in a phenomenological way, an infrared energy scale, namely  $\sqrt{c} $,   in the model.  This action corresponds to a negative cosmological constant $ \Lambda = - \frac{12}{L^2} $ and the  solution considered in the soft wall model has a constant negative curvature $R$. At zero temperature the geometry is  an anti-de Sitter space with radius  $L$. At finite temperature there are two soutions. One is the thermal AdS space and the other the AdS black hole space. One assumes that  the dilaton field does not affect the gravitational dynamics of the theory. This can be interpreted as the assumption that   the fluctuations of $\phi$ are very small as compared to the gravitational ones.   

The thermal AdS space, in the Euclidean signature with compact time direction, is described by
\begin{equation}\label{thermalAdS}
    ds^2= \frac{L^2}{z^2}\left( dt^2 + d\overrightarrow{x}^2 + dz^2 \right)\;,
\end{equation} 
and the AdS with a black hole, by
\begin{equation}\label{BHAdS}
ds^2= \frac{L^2}{z^2}\left( f(z) dt^2 + d\overrightarrow{x}^2 + \frac{dz^2}{f(z)} \right)\;,
\end{equation}
with $f(z) = 1 - z^4/z_h^4$, being $z_h$ the position of the black hole horizon. For the black hole, the time has a period $\beta$ and the temperature is $T = 1/\beta = 1/(\pi z_h)$, to avoid a conical singularity of the metric on the horizon \cite{Hawking:1982dh}. In the thermal AdS case, the periodicity of time is,  in principle, not constrained. However, requiring that the assymtotic limit of the two geometries at $z=\epsilon$, with $ \epsilon \to 0 $,  are  the same, one finds it out  that the period of the thermal AdS is $\beta^\prime = \pi z_h \sqrt{f(\epsilon)}$. 

The AdS geometries  (\ref{thermalAdS}) and (\ref{BHAdS}) are both solutions of the Einstein action,  with curvature $R = -20/L^2$. So that, in both cases, one obtains the on-shell action
\begin{equation}
I_{\text{on-shell}} = \frac{4}{L^2\kappa^2} \int d^5x \sqrt{g}\,e^{-\Phi}\;.
\label{Action01}
\end{equation} 
It is worth to define action  densities  $\mathcal{E}(\epsilon) = I/V$, where $V$ is the trivial transverse tree-dimensional space.  For the thermal AdS it reads
\begin{equation}\label{EAdS}
\mathcal{E}_{AdS}(\epsilon) = \frac{4L^3}{\kappa^2} \int_0^{\beta^\prime} dt \int_\epsilon^\infty dz\, z^{-5} e^{-cz^2}\;, 
\end{equation}
and for the black hole it is:
\begin{equation}\label{EBH}
\mathcal{E}_{BH}(\epsilon) = \frac{4L^3}{\kappa^2} \int_0^{\pi z_h} dt \int_\epsilon^\infty dz\, z^{-5} e^{-cz^2}\;. 
\end{equation}
The regulator $\epsilon$ is necessary due to the presence of ultraviolet divergences in the integration over $z$. However, defining $\bigtriangleup \mathcal{E}$ as the difference between the black hole AdS and the  thermal AdS action densities
\begin{equation}
\bigtriangleup \mathcal{E} = \lim_{\epsilon \rightarrow 0} \left[ \mathcal{E}_{BH}(\epsilon) - \mathcal{E}_{AdS}(\epsilon) \right]\;,
\end{equation}
the UV divergences are cancelled, and one obtains the following result
\begin{equation} \label{deltaEresult}
\bigtriangleup \mathcal{E} = \frac{\pi L^3}{\kappa^2 z_h^3}\left[ e^{-cz_h^2}\left( - 1 + cz_h^2 \right)+ \frac{1}{2} + c^2 z_h^4 \text{Ei}(-cz_h^2)\right]\;,
\end{equation}
where $\text{Ei}(x) = - \int_{-x}^\infty \frac{e^{-t}}{t} dt$. We show in Figure 1 a plot of $ \bigtriangleup \mathcal{E} $ as a function of the temperature. One can see in this figure that for temperatures above a critical value $ T = T_c$, 
$ \, \bigtriangleup \mathcal{E} $ is negative while for smaller values it is positive. So that, there is a  Hawking-Page transition\cite{Hawking:1982dh}. When  $ T > T_c$  the black hole has a smaller action, so it is stable. For lower temperatures, the thermal AdS is the stable phase.  The critical temperature corresponds to  $cz_h^2 = 0.419035$, or, equivalently to:
\begin{equation}
T_c = 0.491720 \sqrt{c}\;. 
\end{equation} 
  
\begin{figure}[!htb]\label{fig1}
	\centering
	\includegraphics[scale=0.55]{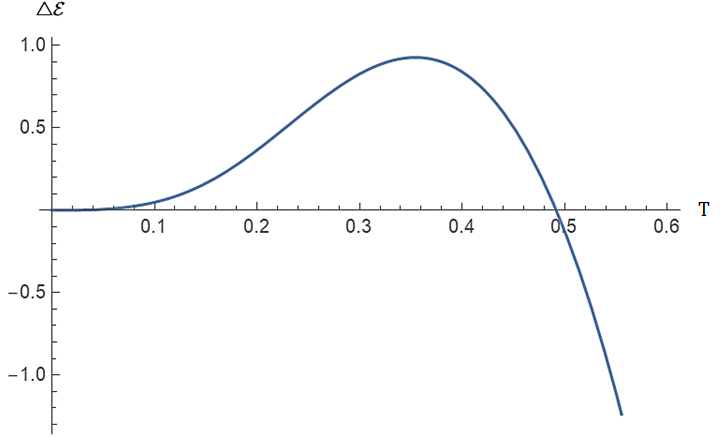}
	\caption{$\bigtriangleup \mathcal{E}$ versus $T$. Action density difference between thermal and black hole AdS spaces, with $L^3/\kappa^2 = c = 1$.}
\end{figure}

If one fixes the parameter $c$ by the spectrum of the  $\rho$-meson, one gets \cite{Herzog:2006ra}  $\sqrt{c}  = 318\, \text{M}eV$. The corresponding deconfinement temperature is $T_c = 191\, \text{M}eV$.  For an interesting alternative study of deconfinement transition in holographic QCD using entanglement entropy see \cite{Dudal:2018ztm}.  

\section{Black hole regularized energy density}

The mass of a black hole  can be determined from the expression \cite{Hawking:1982dh, Witten:1998zw} 
\begin{equation}\label{M}
    M = \frac{\partial I}{\partial \beta}  \;.
\end{equation}
Considering  the regularized difference between the 
black hole and AdS actions 
\begin{equation}\label{Itotal}
I  = I_{BH}  - I_{AdS}\;, 
\end{equation}
  one finds an energy    per unit of transverse volume $V$
\begin{equation}\label{MBH}
 E  = M/V  = \frac{\partial \bigtriangleup \mathcal{E}}{\partial \beta}\;, 
\end{equation}
with $\bigtriangleup \mathcal{E}$ defined by eq. \eqref{deltaEresult}. The natural definition of a black hole energy density $\rho_{BH}(z)$ in the relevant holographic $z$ coordinate  would be one satisfying:
\begin{equation}\label{rhoBHint}
\int_0^\infty \rho_{BH}(z) \, dz =  E \;.  
\end{equation}  
In order to find such a density one can  apply the derivative with respect to  $\beta$ on $  \bigtriangleup \mathcal{E} $ but before integrating over   coordinate $z$. That means, considering $  \bigtriangleup \mathcal{E} $ defined as the difference between the expression in eq. \eqref{EBH}  and the one in  eq. \eqref{EAdS} ,  taking into account the fact that the  limits of integrations depend on $\beta$. 

It is convenient to introduce the coordinate  $u = z/z_h$. Using equations \eqref{EAdS} and \eqref{EBH},
one finds
\begin{eqnarray}\label{deltaE(u)}
 \bigtriangleup \mathcal{E}(u) =  \frac{4L^3}{\kappa^2} \left[ \beta \int_{\epsilon \pi/\beta}^{1} dz\, u^{-5}\left(\frac{\beta}{\pi}\right)^{-4} e^{-c(\frac{u\beta}{\pi})^2} -\left( \beta - \frac{\pi^4\epsilon^4}{2\beta^3}\right) \int_{\epsilon \pi/\beta}^\infty du\, u^{-5} \left(\frac{\beta}{\pi}\right)^{-4} e^{-c(\frac{u\beta}{\pi})^2} \right]\;.
\end{eqnarray}
The derivative of \eqref{deltaE(u)} w.r.t. $\beta$ yields,  after taking the $\epsilon \to 0 $ limit, 
\begin{equation} \label{partialbetaE1a}
E \,=\,  \lim_{\epsilon \to 0} \frac{\partial \bigtriangleup \mathcal{E}}{\partial \beta} = \frac{4L^3}{\kappa^2}\left[ \frac{1}{2z_h^4} + \int_1^\infty du \left( \frac{3 \pi^4}{u^5 \beta^4}+ \frac{2c \pi^2}{u^3 \beta^2}  \right) e^{-c\left(\frac{u\beta}{\pi}\right)^2}  - \frac{7 \epsilon^4}{2} \int_{\frac{\epsilon \pi}{\beta}}^\infty du\,\frac{\pi^8}{u^5 \beta^8}e^{-c\left(\frac{u\beta}{\pi}\right)^2}
            \right]\;.
\end{equation} 
 
The first (non-integrated) term of eq. \eqref{partialbetaE1a}  comes from the differentiation of the integration limit with respect to $\beta$. It can be written in terms of the last term, since one can easily show that
 \begin{equation}\label{1/2zh}
\frac{1}{2z_h^4} =   2 \epsilon^4  \int_{\epsilon\pi/\beta}^\infty du  \frac{\pi^8}{u^5 \beta^8} e^{- c(\frac{u\beta}{\pi})^2}  \;.
\end{equation}
So,  all the terms in the expression \eqref{partialbetaE1a}  for  the energy E can be written as integrals over the coordinate $z$ and   one can associate the integrand with a density. Actually, the result of the  integral (\ref{1/2zh}) does not depend on the value of the upper limit of integration, as long as it is greater than the lower limit. So , we will take it to be at $u = 1$.  Finally, returning to the $z$-variable, one finds 
\begin{equation}
E  = \lim_{\epsilon \to 0}  \left(  \int_{\epsilon}^{z_h} \rho_1^{BH}(z) \,dz + \int_{ z_h}^{ \infty }\rho_2^{BH}(z) \,dz  \right) \;,
\end{equation} 
 \noindent where  
\begin{eqnarray}\label{BHdensity}
\rho^{BH}_1(z) &=& - \frac{4L^3}{\kappa^2} \frac{3 \epsilon^4 }{2 z^5 z_h^4} e^{-cz^2}  \;, \quad\quad \quad  \text{if} \quad \epsilon  \leq z \leq z_h\;, \nonumber\\
\rho^{BH}_2(z) &=& \frac{4L^3}{\kappa^2} \left( \frac{3 }{z^5} +  \frac{2c }{z^3} \right) e^{-cz^2} \;, \quad\, \text{if} \quad z > z_h\;. 
\end{eqnarray} 
 
The expression \eqref{BHdensity} shows that the BH mass is concentrated in the ultraviolet and infrared regions, \textit{i.e.}, in the extremes of $z$-direction. Such a regularized result for the energy density makes it possible to calculate  the corresponding momentum space density, that is necessary in order to find the CE. 

\section{Wien law in the soft wall model }
The BH energy density, $\rho_{BH}(z)$, depends only on the  Poincaré coordinate $z$.  The corresponding Fourier transform takes the form
\begin{eqnarray}\label{rho(k)}
 \widetilde{\rho}(k) = \frac{1}{2\pi} \lim_{\epsilon \to 0} \left( \int_\epsilon^{z_h} dz\, \rho^{BH}_1(z) e^{i k z} + \int_{z_h}^\infty dz\, \rho^{BH}_2(z) e^{i k z} \right)\;,
\end{eqnarray} 
with $\rho^{BH}_1(z)$ and $\rho^{BH}_2(z)$  defined by eq. \eqref{BHdensity}.  The relevant quantity for the calculation of the modal  fraction, and then the CE,  is the   squared absolute value of the black hole energy density in Fourier space, that can be written as:  
\begin{eqnarray}
    \vert \widetilde{\rho}(k) \vert^2 &=& \left[ \frac{1}{2\pi} \lim_{\epsilon \to 0} \left(\int_\epsilon^{z_h} \rho^{BH}_1(z) \cos(kz)\,dz + \int_{z_h}^\infty \rho^{BH}_2(z) \cos(kz)\,dz \right)\right]^2 \nonumber\\ 
&+& \left[ \frac{1}{2\pi} \lim_{\epsilon \to 0}  \left(\int_\epsilon^{z_h} \rho^{BH}_1(z) \sin(kz)\,dz + \int_{z_h}^\infty \rho^{BH}_2(z) \sin(kz)\,dz \right)\right]^2 \;.
\label{quadraticfourier}
\end{eqnarray}

One can use numerical methods in order to compute these integrals, since they do not present analytical solutions. The results must be independent of the UV regulator $\epsilon$. As the numerical computation is performed with a finite value of $ \epsilon $, the limit $ \epsilon  \to 0 $  is effectively obtained by identifying the order of magnitude for which taking smaller values of $\epsilon$ would  not change the results, considering  the degree of precision in which one is working.  We found it out that from $\epsilon \sim  10^{-16}$ to smaller values there is no change in any of the results of this work, so this value was used in our computations. 

The functions $\vert \widetilde{\rho}(k) \vert^2$   present a global maximun   $\vert \widetilde{\rho}(k) \vert^2_{max}$, at a finite value of $ k$,  that we will name as $k_{max}$. Analysing the variation of the value of $k_{max}$ with the temperature, it is  found  that 
\begin{equation}\label{kT}
\frac{k_{max}}{T} = K, 
\end{equation}  
where $ K $ is a constant. Using the normalization $\frac{4L^3}{2\pi \kappa^2}=1$  and taking $c=1$, that corresponds to working with the dimensionless temperature $T \equiv T/\sqrt{c}$, one finds that  $K = 8.790 \times 10^4$, with an error of $\pm 0.001 \times 10^4$ or $0.01\%$.   In Figure 2-(A), we plot the absolute value of the BH energy density at $T=0.65$, which shows a behaviour that is the same for all values of $T$. Namely,  $\vert \widetilde{\rho}(k) \vert^2$ goes  to zero as $k \rightarrow 0$, for $ k > 0$ it  increases until it reaches the global maximum, and then oscillates, tending to zero for large momentum.  
\begin{figure}[!htb]\label{figWien1}
	\centering
	\includegraphics[scale=0.53]{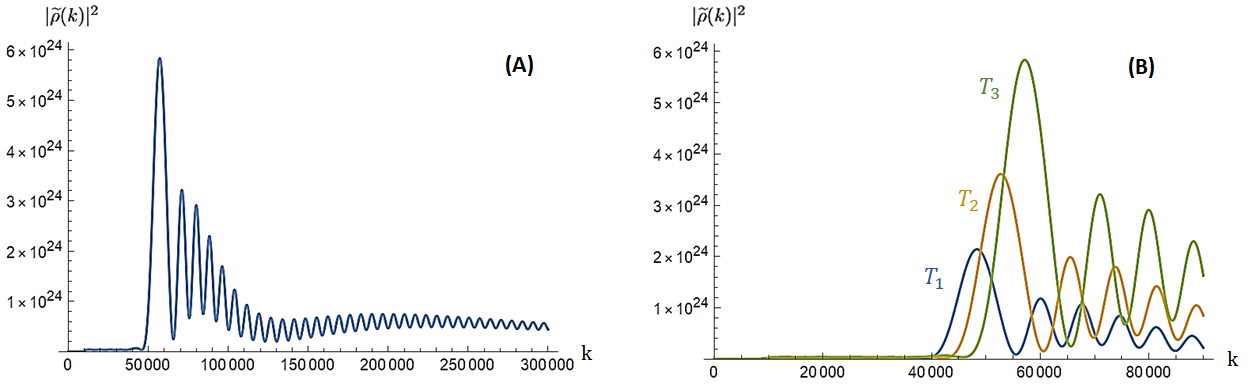}
	\caption{(A) $\vert \widetilde{\rho}(k) \vert^2$ versus momentum at $T=0.65$; (B) the same at $T1 = 0.55$ (blue), $T2 = 0.6$ (orange); and $T3 = 0.65$ (green).}
\end{figure}
As $T$ increases,  $\vert \widetilde{\rho}(k) \vert^2_{max}$ and $k_{max}$  increase, as shown in Figure 2-(B), where we plot $\vert \widetilde{\rho}(k) \vert^2$ at three different temperatures, $T_3>T_2>T_1$. 
 
 This behaviour is represented in Figure 3, where we plot the points displayed in Table 1.For the confined phase, corresponding to $T < T_c $, there is no black hole, so there is  no energy density.  For $T>T_c$, the points represent the  relation described by eq. \eqref{kT}.
\begin{table}[h]
\centering
\begin{tabular}[c]{|c|c|c||c|c|c|}
\hline 
 $T$ & $\vert \widetilde{\rho}_{BH}(k) \vert^2_{max}$ & $k_{max}$   & $T$ & $\vert \widetilde{\rho}_{BH}(k) \vert^2_{max}$ & $k_{max}$  \\
\hline
$\,\,\,0.50\,\,\,$ & $\,\,\,1.2092\times 10^{24}\,\,\,$ & $\,\,\,43949\,\,\, $ & $\,\,\,1.00\,\,\,$ &$\,\,\,7.7388\times 10^{25}\,\,\,$ & $ \,\,\, 87900\,\,\, $ \\
\hline
$\,\,\,0.55\,\,\,$ & $\,\,\,2.1422\times 10^{24}\,\,\,$ & $\,\,\, 48345\,\,\, $ & $\,\,\,1.05\,\,\,$ & $\,\,\,1.0371\times 10^{26}\,\,\,$ &$ \,\,\,92295\,\,\,  $ \\
\hline 
$\,\,\,0.60\,\,\,$ &$\,\,\,3.6106\times 10^{24}\,\,\,$ & $  52740 $    & $\,\,\,1.10\,\,\,$ &$\,\,\,1.3710\times 10^{26}\,\,\,$ &$96683 $\\
\hline
$\,\,\,0.65\,\,\,$ &$\,\,\,5.8365\times 10^{24}\,\,\,$ & $57135 $  & $\,\,\,1.15\,\,\,$ &$\,\,\, 1.7900\times 10^{26}\,\,\,$ & $ 101088 $   \\
\hline
$\,\,\,0.70\,\,\,$ &$\,\,\,9.1046\times 10^{24}\,\,\,$ &$ 61530 $  & $\,\,\,1.20\,\,\,$ &$\,\,\,2.3108\times 10^{26}\,\,\,$ & $ 105474 $  \\ 
\hline
$\,\,\,0.75\,\,\,$ &$\,\,\,1.3773\times 10^{25}\,\,\,$ &$ 65924 $    & $\,\,\,1.25\,\,\,$ &$\,\,\,2.9521\times 10^{26}\,\,\,$ & $ 109873  $\\
\hline 
$\,\,\,0.80\,\,\,$ &$\,\,\,2.0287\times 10^{25}\,\,\,$ &$   70321 $   & $\,\,\,1.30\,\,\,$ &$\,\,\,3.7152\times 10^{26}\,\,\,$ & $ 114275  $\\
\hline
$\,\,\,0.85\,\,\,$ &$\,\,\, 2.9187\times 10^{25}\,\,\,$ &$ 74712 $  &  $\,\,\,1.35\,\,\,$ &$\,\,\,4.6593\times 10^{26}\,\,\,$ &$  118665 $    \\
\hline
$\,\,\,0.90\,\,\,$ &$\,\,\,4.1127\times 10^{25}\,\,\,$ & $79117 $ & $\,\,\,1.40\,\,\,$ &$\,\,\, 5.7955\times 10^{26}\,\,\,$ & $ 123066 $   \\ 
\hline
$\,\,\,0.95\,\,\,$ & $\,\,\,5.6887\times 10^{25}\,\,\,$ & $ 83505 $    & $ \,\,\,1.45\,\,\,$ &$\,\,\,7.1537\times 10^{26}\,\,\,$ &$ 127457$ \\
\hline 
\end{tabular}   
\caption{Maximum of BH density, $\vert \widetilde{\rho}(k) \vert^2_{max}$, with the corresponding momentum $k_{max}$, at different temperatures.}
\label{table1}
\end{table} 
\begin{figure}[!htb]\label{fig2}
	\centering
	\includegraphics[scale=0.55]{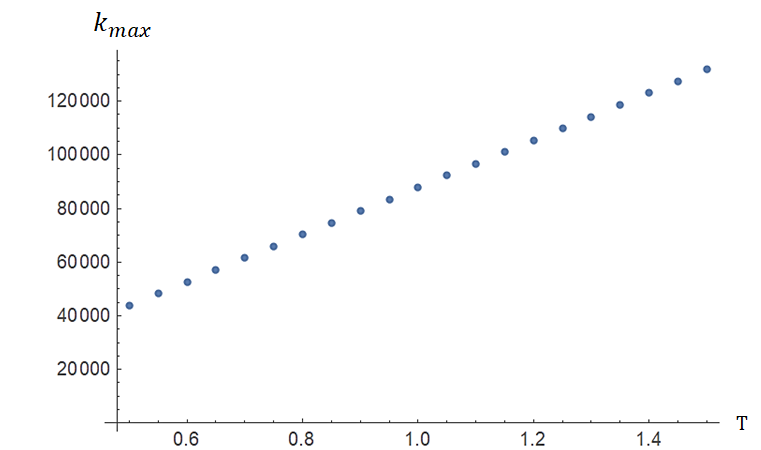}
	\caption{$k_{max}$ versus $T$ for $T > T_c$.}
\end{figure}

The  wavelength associated with the momentum $k_{max} $ is  $\lambda_{max} =  2\pi/k_{max}$. Replacing this relation in  \eqref{kT}, one gets the analogous of Wien's displacement law
\begin{equation}\label{Wien}
\lambda_{max} T = C\;, 
\end{equation}
wherein $C=\frac{2\pi}{K} = 7.148 \times 10^3$. This equation shows that the BH energy density will peak at different wavelength that are inversely proportional to the temperature, analogously to the black-body radiation spectrum.

\section{Configuration entropy and stability}

Now, let us study the stability of the soft wall model  by computing the configuration entropy (CE)   following the prescriptions presented in the previous sections.  
First we calculate the modal fraction \eqref{modalfraction}, as defined in the previous section, being $k_{max}$ the value of $k$ where $ \vert \widetilde{\rho}(k) \vert^2$ assumes the maximum value. The square of the Fourier transform of the energy density is obtained from equation (\ref {quadraticfourier}) using  the spatial density defined in eq. (\ref{BHdensity}).   Then the CE is calculated using equation \eqref{positionentropy}. This way, one can determine the CE as a function of the temperature. Below $T_c = 0.491720$,  the stable phase is the thermal AdS. So, there is no black hole and the CE is   zero since,  as expressed in  \eqref{Itotal}, we are considering the AdS as the  ``empty space''     background.  In the deconfined phase, where the black holes are stable, the numerical computations of the CEs at the same temperatures of Table 1, lead to the results displayed in Table 2 and ploted in Figure 4.    
\begin{table}[h]
\centering
\begin{tabular}[c]{|c|c||c|c|}
\hline 
 $T$ & Black hole CE  & $T$ & Black hole CE   \\
\hline
$\,\,\,0.50\,\,\,$ & $327276 $ & $\,\,\,1.00\,\,\,$ & $  654774 $ \\
\hline
$\,\,\,0.55\,\,\,$ & $360026 $ & $\,\,\,1.05\,\,\,$ & $  687282 $ \\
\hline 
$\,\,\,0.60\,\,\,$ & $392764 $ & $\,\,\,1.10\,\,\,$ & $  720018 $ \\
\hline
$\,\,\,0.65\,\,\,$ & $425406 $ & $\,\,\,1.15\,\,\,$ & $  752790 $ \\
\hline
$\,\,\,0.70\,\,\,$ & $458194 $ & $\,\,\,1.20\,\,\,$ & $  785476 $ \\
\hline
$\,\,\,0.75\,\,\,$ & $490936 $ & $\,\,\,1.25\,\,\,$ & $  818200 $ \\
\hline 
$\,\,\,0.80\,\,\,$ & $523682 $ & $\,\,\,1.30\,\,\,$ & $  850874 $ \\
\hline
$\,\,\,0.85\,\,\,$ & $556308 $ & $\,\,\,1.35\,\,\,$ & $  883674 $ \\
\hline
$\,\,\,0.90\,\,\,$ & $589096 $ & $\,\,\,1.40\,\,\,$ & $  916450 $ \\
\hline
$\,\,\,0.95\,\,\,$ & $621846 $ & $\,\,\,1.45\,\,\,$ & $  949192 $ \\
\hline 
\end{tabular}   
\caption{Black hole CE at different temperatures.}
\label{table1}
\end{table} 
   
\begin{figure}[!htb]\label{figCE}
	\centering
	\includegraphics[scale=0.55]{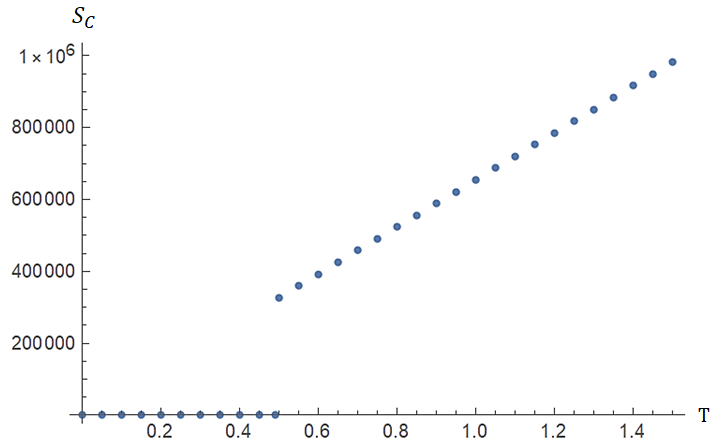}
	\caption{CE versus T.}
\end{figure}

One can see in Figure 4 that the CE varies linearly with the temperature.  In particular, it is found that 
\begin{equation}\label{S_C}
S_C = K_cT\;,
\end{equation}
where $K_c = (6.546 \pm 0.001 )  \times 10^5$, meaning that black holes become more unstable at higher temperatures. Such a result is consistent with the expectation that   black holes are subject to a process of evaporation by Hawking  radiation, and this effect should increase with the temperature.    

Another outcome of this result for the CE is that there is also a linear relation between the 
CE and   $k_{max} $:   $S_C \sim k_{max}$.  
The jump from $S_C = 0$ in the confined hadronic phase to $S_C \sim T$ in the plasma phase, represents in a consistent way the release of the color degrees of freedom that happens at $T=T_c$. From the correlation between the Wien law and the CE, one concludes that states with smaller values of $k_{max}$   are more stable.

\section{Analysis of the results and conclusions}  

Studying the thermodynamics of the black hole geometry  in the soft wall AdS/QCD model, we obtained, in eq. (\ref  {BHdensity}),   a regularized form for the energy density in the holographic coordinate $z$. This regularized expression made it possible the calculation of the square of the absolute value of energy density in momentum space 
 $\vert \widetilde{\rho}(k) \vert^2$, given by eq. \eqref{quadraticfourier}. The results obtained for the black hole phase display a behaviour with the form of the  Wien's displacement law for black body radiation. Associating a  wavelength  $ \lambda$ with 
 the inverse of $k$ one finds it out that the the value $k_{max} $ where the maximum of $\vert \widetilde{\rho}(k) \vert^2$ occurs, corresponds to a wavelength $ \lambda_{max} $ that is inversely proportional to the temperature,  as shown  in eq. \eqref{Wien}.
  
 We also found  that the CE is  proportional to $k_{max}$ and to the temperature. This  is a nontrivial issue. Recently  it was show in Ref. \cite{Braga:2020opg}   that the BH configuration entropy in the hard wall holographic model is proportional to $\log{T}$, showing a different behaviour with respect to the one found here for the soft wall. It is known that the soft wall presents a better holographic description of hadronic matter since the mass spectra obtained with this model present linear Regge trajectories, as opposed to the hard wall model. So, it is interesting to see that it also present a richer structure from the point of view of the configuration entropy. The gap between the thermal AdS and black hole   phases is consistent with the transition from hadronic to   the  plasma  phase, or confinement/deconfinement transition. This corresponds to the liberation of the color degrees of freedom at the Hawking-Page critical temperature. 

The result that the CE increases with the temperature, indicating that the black hole becomes more unstable, is  consistent with the expectation that black holes undergo  a process of evaporation that increases in intensity  with the  temperature.   The evaporation of anti-de Sitter black holes with planar horizon was discussed in Ref.  \cite{Ong:2015fha}. There, it was shown that the rate of change in the mass is proportional to a positive power of the mass itself. So, the rate of change of the mass increases with the temperature. The conclusion that follows is that the CE captures consistently the dynamical instability of the black hole phase, associated with the evaporation.  In the gauge theory side of the holographic duality, this instability could be 
associated with the instability of the plasma, that has a finite lifetime, undergoing a process of hadronization after this time. Another interesting point to be remarked is that the vanishing of the CE  for $T < T_c$ is a result consistent with the interpretation that the instability captured by the CE is associated with the black hole evaporation. In the confined phase there is no black hole, so there is no evaporation.

\noindent \textbf{Acknowledgments}: The authors are supported by  CNPq - Conselho Nacional de Desenvolvimento Cient\'ifico e Tecnol\'ogico. This work received also support from  Coordena\c c\~ao de Aperfei\c coamento de Pessoal de N\'ivel Superior - Brasil (CAPES) - Finance Code 001.

\end{document}